\documentclass[pra, superscriptaddress, twocolumn, showpacs]{revtex4}

\usepackage{amsfonts}
\usepackage{amssymb}
\usepackage[dvips]{graphicx}
\usepackage{amsmath}

\newcommand{\bra}[1]    {\langle #1|}
\newcommand{\ket}[1]    {| #1 \rangle}

\begin{document}
\title{Experimental quantum-enhanced estimation of a lossy phase shift}
\author{M.~Kacprowicz}
\affiliation{Institute of Physics, Nicolaus Copernicus University, ul.~Grudziadzka 5, PL-87-100 Toru\'{n}, Poland}
\author{R.~Demkowicz-Dobrza{\'n}ski}
\affiliation{Institute of Physics, Nicolaus Copernicus University, ul.~Grudziadzka 5, PL-87-100 Toru\'{n}, Poland}
\author{W.~Wasilewski}
\affiliation{Institute of Experimental Physics, University of Warsaw, Ho\.{z}a 69, PL-00-681 Warsaw, Poland}
\author{K.~Banaszek}
\affiliation{Institute of Physics, Nicolaus Copernicus University, ul.~Grudziadzka 5, PL-87-100 Toru\'{n}, Poland}
\author{I.~A.~Walmsley}
\affiliation{Clarendon Laboratory, University of Oxford, Parks Road, Oxford OX1 3PU, United Kingdom}

\date{\today}

\pacs{03.65.Ta, 06.20.Dk, 42.50.Lc, 42.50.St}
\maketitle
{\bf
A paradigm for quantum-enhanced precision metrology is found in optical interferometry
\cite{InterferometryReview}, capable to sense physical quantities as diverse as position, time delay or temperature through a measurement of a phase shift.
When standard light sources are employed, the precision of the phase determination is limited by the shot noise, whose origin can be traced to the random manner in which individual photons emerge from the interferometer.
Quantum entanglement provides means to exceed this limit
\cite{CavePRD81,GranSlusPRL87,XiaoWuPRL87,HollBurnPRL93,SandMilbPRL95} with the celebrated example of N00N states \cite{BollItanPRA96,DowlPRA98,WaltPanNAT04,MitcLundNAT04} that saturate the ultimate Heisenberg limit on precision
\cite{GiovLloySCI04}, but at the same time are extremely fragile to losses \cite{HuelMaccPRL97,RubiKausPRA07,GilbHamrJOSAB08}. In contrast, we provide experimental evidence that appropriately engineered quantum states \cite{DornDemkPRL09} outperform both standard and N00N states in the precision of phase estimation when losses are present. This shows that the quantum enhancement of metrology is possible even when decoherence is present, and that the strategy for realising the enhancement is quite distinct from protecting quantum information encoded in light \cite{BourEiblPRL05,LuGaoPNAS08}.
}
\begin{figure}[t]
\includegraphics[width=0.42 \textwidth]{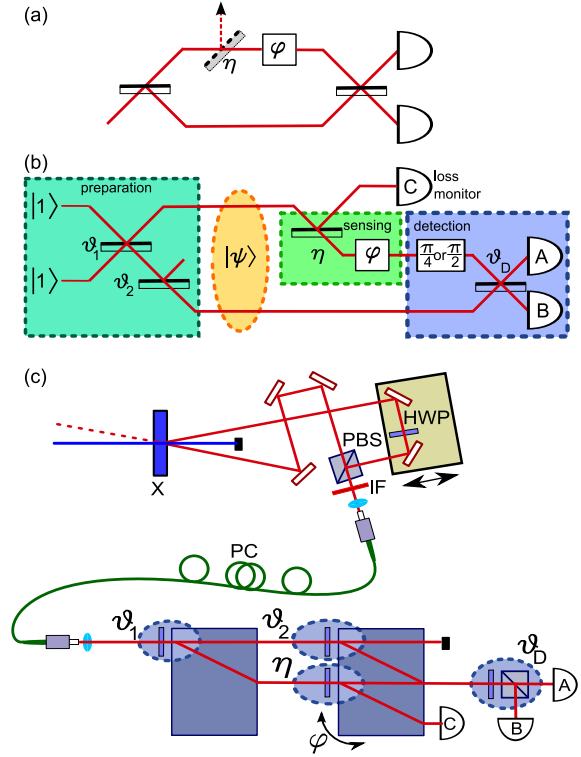}
\caption{(a) The generic Mach-Zennder interferometer fed with a classical light beam. The phase shift $\varphi$ inside the interferometer modulates the output intensities. Possible losses accompanying the phase shift are represented by a beam splitter with power transmission $\eta$;
(b) Linear network to demonstrate optimal phase estimation. The beam splitters with power transmission $\vartheta_1$ and $\vartheta_2$ prepare the optimal probe state $\ket{\psi}$. Phase information is read out by applying a conditional phase shift $\frac{\pi}{4}$ or $\frac{\pi}{2}$ and interfering the paths on a beam splitter with a suitably chosen transmission $\vartheta_D$; (c) Experimental realization of the linear network. The input to the setup are either ultraviolet 390~nm (blue solid line) or fundamental 780~nm (dashed red line) pulses. Ellipses mark locations of effective beam splitters acting on orthogonal polarizations. For details, see Methods.}
\label{Fig:Interferometry}
\end{figure}

The Mach-Zehnder interferometer shown in Fig.~\ref{Fig:Interferometry}(a) serves to illustrate the principle of phase estimation. The input light is divided into two beams and one of them probes the phase shifting element. Recombining the two beams maps the acquired phase information onto the output intensity, read out via photodetection. The photodetection signal exhibits fluctuations which result in the shot noise limiting the measurement precision. In the semiclassical picture this effect is attributed to the uncertainty in the photoelectron number
generated by an incident beam even if its intensity is constant \cite{SemiclassicalPhotodetection}. In the fully quantum picture, light is composed of elementary quanta---photons---whose behaviour can be predicted only statistically, hence leading to fluctuations at the interferometer output
\cite{CavePRD81}. The shot noise scales as $1/\sqrt{N}$, where $N$
is the average number of photons used for the measurement. For a lossy phase shift, it is necessary to readjust the splitting ratio for the input light which yields a generalized bound on precision \cite{DemkDornXXX09} in the form of the \emph{standard interferometric limit} (SIL).

More generally, the two beams---one sent through the sample and the second one serving as the reference---can be prepared in a certain probe state $\ket{\psi}$. According to quantum estimation theory \cite{Helstrom1976}, the precision
$\delta\varphi$ of the estimate is restricted by the quantum Cram\'{e}r-Rao bound $\delta\varphi \ge 1/\sqrt{F}$, where $F$ is the quantum Fisher information. In the case of a phase measurement, $F$ is proportional to the variance of the photon number $\hat{n}_s$ sent through the sample:
\begin{equation}
\label{eq:cr}
F \bigl( \ket{\psi} \bigr)=4[\bra{\psi} \hat{n}^2_s \ket{\psi} -
(\bra{\psi} \hat{n}_s \ket{\psi})^2].
\end{equation}
Thus phase sensitivity is intimately related to the uncertainty of the photon number in the sensing arm.
If the total number of photons is limited to $N$, this variance is maximized by a N00N state \cite{BollItanPRA96,DowlPRA98} of the form
$\bigl( \ket{N0} - \ket{0N} \bigr)/\sqrt{2}$, describing a coherent, equally weighted superposition of all the photons traveling in either the sensing or the reference arm of the interferometer. Here kets enclose a pair of figures that specify the numbers of photons in the two interferometer arms. An application of N00N states improves the precision to  $1/N$, known as the Heisenberg limit \cite{GiovLloySCI04}, which  defines the ultimate lower bound for phase estimation \cite{BrauLanePRL92}.
  This advantage however fades away if the phase shift is accompanied by attenuation: even loss of only one photon destroys completely the superposition and the associated capability to acquire phase information
\cite{RubiKausPRA07,GilbHamrJOSAB08}.

\begin{figure}
\includegraphics[width=0.52\textwidth]{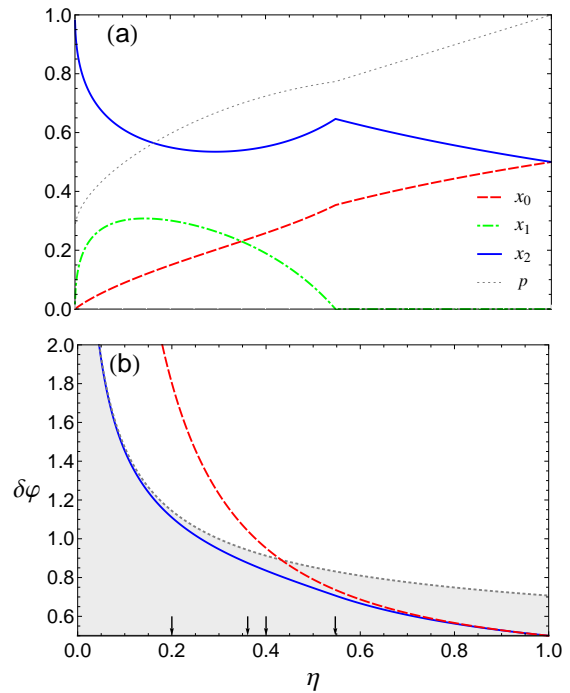}
\vspace{-1cm}
\caption{ (a) Weights $x_2$, $x_1$, and $x_0$ of components in the optimal two-photon probe state defined in
Eq.~(\protect\ref{eq:genstate}) as a function of the phase shift transmission $\eta$, shown together with the probability $p$ of preparing the optimal state using the scheme shown in Fig.~\protect\ref{Fig:Interferometry}(b) based on two beam splitters; (b) the precision of phase estimation using two-photon optimal states (solid blue), N00N states (dashed red), and the standard interferometric limit (dotted gray). The shaded area depicts the non-classical region, and the four arrows mark transmissions $\eta$ for which experimental data have been taken.}
\label{Fig:OptimalStatesandResolution}
\end{figure}

The simplest non-trivial example of a N00N state is a superposition $(\ket{20}- \ket{02})/\sqrt{2}$ which can be prepared using the Hong-Ou-Mandel effect \cite{HongOuPRL87}, by sending two photons onto a balanced beam splitter. After transmission through the phase element the two superposition terms acquire a relative phase $2\varphi$, leading to a doubled density of the interference fringes  \cite{RariTapsPRL90} which is behind the precision enhancement. When a photon is lost in the sample, the quantum state of light inside the interferometer collapses to $\ket{10}$, which does not have any phase sensitivity. In order to avoid this problem, we take the probe state in the form of a three-component superposition
\begin{equation}
\label{eq:genstate}
\ket{\psi}= \sqrt{x_2} \ket{20}+\sqrt{x_1}\ket{11} - \sqrt{x_0} \ket{02},
\end{equation}
where the non-negative weights $x_0$, $x_1$, and $x_2$ sum up to one. Let $\eta$ be the intensity transmission of the sample.
We will denote by $p_l$ the probability of losing $l$ photons and write the corresponding conditional state
as  $\sqrt{p_l}\ket{\psi_l}$, where the ket $\ket{\psi_l}$ is normalized.
When zero or one photon is lost, the conditional states read:
\begin{eqnarray}
\sqrt{p_0}\ket{\psi_0} & = & \eta \sqrt{x_2} \ket{20}+\sqrt{\eta x_1} \ket{11} -   \sqrt{x_0} \ket{02}
\nonumber \\
\sqrt{p_1}\ket{\psi_1} & = & \sqrt{2 \eta (1-\eta) x_2}\ket{10} + \sqrt{(1-\eta) x_1} \ket{01}
\end{eqnarray}
while only for the loss of both the photons the phase sensitivity is destroyed. Because the conditional states are distinguishable by the total photon number, the quantum Fisher information is given by a combination $p_0 F \bigl( \ket{\psi_0} \bigr)
+ p_1 F \bigl( \ket{\psi_1} \bigr)$ \cite{DemkDornXXX09}. Maximization of this quantity yields optimal values of
$x_0$, $x_1$, and $x_2$, depicted in Fig.~\ref{Fig:OptimalStatesandResolution}(a). The resulting precision is compared with the SIL and that achievable with original two-photon N00N states in Fig.~\ref{Fig:OptimalStatesandResolution}(b), clearly showing preservation of the non-classical enhancement.

In order to demonstrate the robustness of the optimal probe states we built a passive linear-optics network shown schematically in Fig.~\ref{Fig:Interferometry}(b), fed with photon pairs $\ket{11}$ produced via spontaneous parametric down-conversion \cite{BurnWeinPRL70}. Preparation of the optimal state is accomplished by sending a pair of photons to a beam splitter characterized by an intensity transmission $\vartheta_1$, which produces a state $\sqrt{2 \vartheta_1 (1-\vartheta_1)} \bigl( \ket{20} - \ket{02} \bigr) + (2\vartheta_1 -1) \ket{11}$. In particular, taking $\vartheta_1=\frac{1}{2}$ yields the two-photon N00N state. In order to change the weights between the components $\ket{20}$ and $\ket{02}$, a second beam splitter with an intensity transmission $\vartheta_2$ is inserted into one of the emerging beams. Provided that both the photons are transmitted, the prepared state has the unnormalized form
$
\sqrt{2 \vartheta_1 (1-\vartheta_1)} \ket{20} + \sqrt{\vartheta_2}(2\vartheta_1 -1) \ket{11} - \vartheta_2 \sqrt{2 \vartheta_1(1-\vartheta_1)} \ket{02}
$.
A suitable choice of $\vartheta_1$ and $\vartheta_2$ allows us to prepare optimal states with the success rate shown in Fig.~\ref{Fig:OptimalStatesandResolution}(a).
Next, one of the produced beams undergoes a phase shift $\varphi$ accompanied by a controlled loss introduced by the non-unit transmission $\eta$ of a beam splitter. Lost photons are monitored with a photon counter C. The attenuated state enters the detection stage, which is designed to emulate a measurement saturating the Cram\'{e}r-Rao bound at $\varphi=0$, thus optimally extracting the encoded phase shift. The two beams are interfered on a beam splitter with an intensity transmission $\vartheta_D$, followed by photon counters A and B. If the probe state does not include the $\ket{11}$ component, or only one photon makes it to the detection stage, we take $\vartheta_D = \frac{1}{2}$, while if neither of these two conditions holds, the beam splitter transmission needs to be adjusted for the specific value of losses. Moreover,
reaching the Cram\'{e}r-Rao bound requires a conditional phase shift before the beams are brought to interfere: $\frac{\pi}{4}$ if no photon is lost and $\frac{\pi}{2}$ if only one photon makes it through the interferometer. This adjustment shifts the steepest slopes of the coincidence fringes to the vicinity of $\varphi=0$. The adaptive nature of the optimal measurement requires in principle a joint quantum nondemolition photon number measurement
\cite{QND} on both the beams after probing the phase shift, and then adjusting the beam splitter characteristics. In a passive linear optics implementation, this may be simulated by collecting separate data for two settings and postselecting events according to the number of photons registered by the photon counter C.
The limitations of the state preparation procedure, together with mode coupling and photon counting efficiencies mean that the measurements must be carried out in the coincidence basis \cite{RalpLangPRA02}. The data used as the input for the estimation procedure are two-fold counting events for all the combinations of the photon counters.

\begin{figure}
\includegraphics[width=0.47\textwidth]{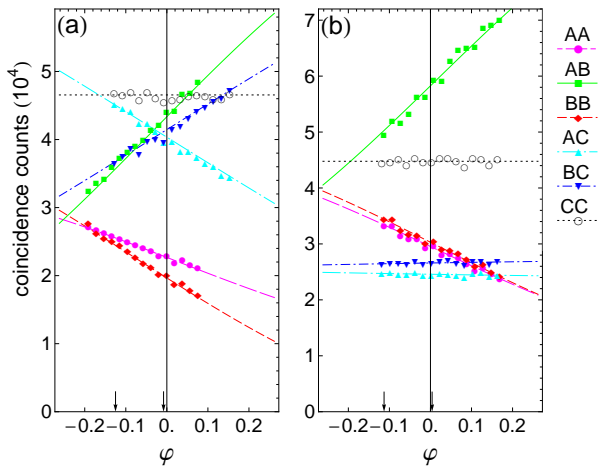}
\caption{Two-fold coincidence rates measured with photons prepared in (a) the optimal state and (b) the N00N state for the transmission $\eta=0.361$ and 15 phases around $\varphi=0$, taken with either $\frac{\pi}{2}$ (coincidences AC, BC, CC) or $\frac{\pi}{4}$ (coincidences AA, BB, AB) additional phase
shift before the beams interfere. The curves depict fitting of a model including experimental imperfections. Vertical arrows mark phase values used to obtain histograms in Fig.~\protect\ref{Fig:Histograms}.}
\label{Fig:Fringes361}
\end{figure}
We implemented the linear optics network presented above using a pair of calcite displacers
\cite{OBriPrydNAT03}, as described in Methods.
The two-fold coincidences are labeled with the paths taken by the photons at the output of the setup. Their dependence on the phase shift $\varphi$ around the point of operation is illustrated in Fig.~\ref{Fig:Fringes361} for optimal and N00N states. The measurements were taken for four different values of loss corresponding to beam splitter transmissions of $\eta = 0.2, 0.361, 0.4$, and $0.547$, employing two-photon optimal and N00N states, as well as using attenuated laser pulses to establish the SIL. For each value of losses, we performed first scans over the entire $2\pi$ range of the phase shift $\varphi$ in order to calibrate the apparatus. The phase dependence of the coincidences counts was used to fit the parameters of the theoretical model \cite{RafalsModel}, which includes the imperfections of the state preparation and linear-optics manipulations. Next, precise measurements were carried out for $15$ values of the phase around the most sensitive point $\varphi=0$, with an increment of $0.02$~rad. In order to acquire information about the estimation uncertainty, for each of the selected phases we took between 100 and 300 series of measurements, one series comprising on average 2000 coincidence events.

\begin{figure}
\includegraphics[width=0.43\textwidth]{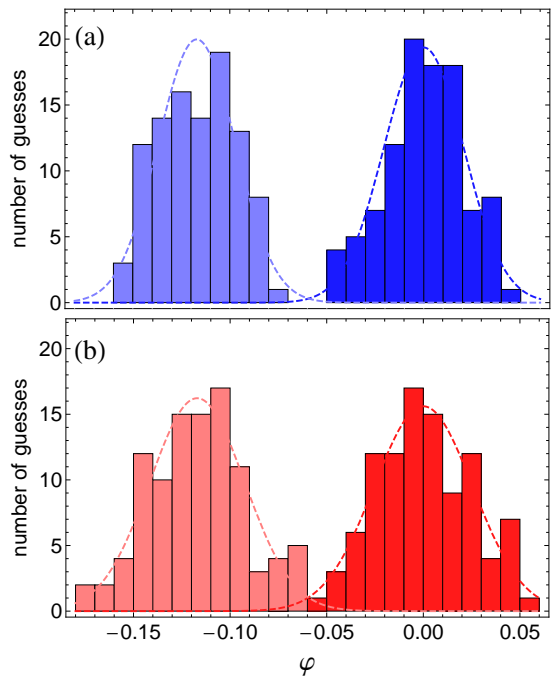}
\caption{Histograms of phase estimates for $\eta=0.361$ obtained using photons prepared in (a) the optimal state and (b) the N00N state, probing two phase shifts differing by 0.12~rad. Each phase estimate was obtained from a series comprising approx.\ $2000$ coincidences.}
\label{Fig:Histograms}
\end{figure}
\begin{figure}
\includegraphics[width=0.48\textwidth]{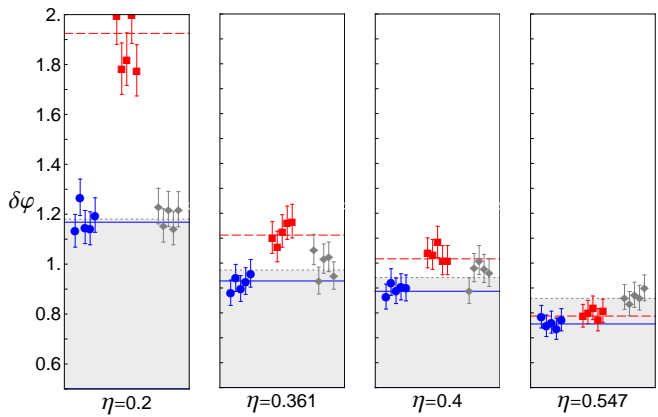}
\caption{Uncertainties of phase estimates obtained using two-photon optimal (circles) and N00N (squares) states, as well as attenuated laser pulses in the SIL regime (diamonds), rescaled by the square root of the number of coincidences. For each transmission $\eta$, data are shown for five phases $\varphi= 0,\pm 0.2,\pm 0.4$~rad. Horizontal lines represent the theoretical Cram{e\'r}-Rao bounds for given classes of input states taking into account imperfections
of the interferometer.}
\label{Fig:StandardDev}
\end{figure}
A phase estimate was extracted from each series using the maximum-likelihood method \cite{MaxLik}, which is known to saturate the Cram\'{e}r-Rao bound when applied to large samples. Histograms of phase estimates for two phases separated by $0.12$~rad, obtained using experimentally generated optimal and N00N states, are shown in Fig.~\ref{Fig:Histograms} along with fitted gaussian distributions. In order to gain a quantitative insight into the improvement, we depict in Fig.~\ref{Fig:StandardDev} the estimation uncertainty for five phases closest to $\varphi=0$, when the sensitivity of the measurement scheme is highest.
The uncertainties obtained from experimental data
are rescaled by the square root of the number of registered coincidences in order to
obtain the figure of merit of the {\em  effective uncertainty per photon pair} that can be compared with the theoretical Cram\'{er}-Rao bound.
The advantage of using the optimal states is clearly seen for all the four values of losses used in the measurements. The results match well the theoretical model including experimental imperfections.

%%Measurements are taken with the optimal and N00N two-photon states, as well as for attenuated laser light used in the SIL.
Quantum-enhanced metrology requires new strategies to cope with decoherence, which are intrinsically different from those targeting protection of optical qubits for quantum information processing applications \cite{BourEiblPRL05,LuGaoPNAS08}. Our experiment demonstrates the fundamental feasibility of such strategies.
Further developments, such as exploiting multiphoton states
\cite{WaltPanNAT04,MitcLundNAT04} and going beyond the coincidence basis \cite{DeterministicSources}, will allow a proper assessment of the capabilities of such approaches including all resources.

\section*{Methods}

Photon pairs are generated in a 1~mm long bismuth borate crystal X oriented for type-I process, pumped with a doubled output from a Ti:sapphire oscillator emitting a 80~MHz train of a 780~nm central wavelength and 140~fs duration pulses. The photons, collected from the opposite ends of a cone with the half-opening angle of 4$^\circ$ are converted into orthogonal polarizations using a half-wave plate HWP, combined on a polarizing beam splitter PBS, filtered through a 5~nm interference filter IF, and coupled into a single-mode fibre. The contribution of higher order photon numbers is negligible. The relative delay between the photons is controlled by a motorized trombone placed in the path of one of the photons. The fibre is wound on a manual polarization controller PC to recover the $\ket{11}$ state at the output of the fibre. We found an adequate description for the the two-photon state at the output of the fibre of the form $\sqrt{1-\epsilon} \ket{11} + e^{i\delta} \sqrt{\epsilon/2} \bigl( \ket{20} + \ket{02} \bigr)$, where $\epsilon$ is typically of the order of $0.05\%$, and the phase $\delta$ is introduced by the fibre. The second term can be attributed to a combination of polarization-dependent loss and birefringence in the single mode fibre delivering photon pairs to the interferometer. Its presence is seen in
phase scans for N00N states and non-zero losses shown in Fig.~\ref{Fig:Fringes361}(b) as a faint modulation of coincidences AC and BC, which in principle should remain constant. The setup also enables to launch into the fibre 780~nm fundamental-wavelength laser pulses to determine the SIL.

The actual implementation of the linear optics network, shown in Fig.~\ref{Fig:Interferometry}(c), is based on a pair of calcite displacers and a Glan-Taylor polarizer, which ensures high visibility and stability of interference. The beam splitters are realized in a common-path version operating on two orthogonal polarizations, with the splitting ratios defined by the orientation of half-wave plates located before the birefringent medium. The second displacer is mounted on a rotation stage with the axis perpendicular to the plane of the setup. The orientation of the displacer defines the phase shift to be determined. The photons leaving the setup are coupled into 50:50 multimode fibre couplers, whose ends are plugged into single photon counting modules. Coupling into multimode fibres is adjusted to equalize effective efficiencies of the modules. In this arrangement, only 50\% of events AA, BB, and CC are detected. In order to compensate for this inefficiency, in the numerical postprocessing half of events AB, AC, and BC has been randomly removed.

The indistinguishability of the interfering two-photon pathways has been tested by setting $\vartheta_1=\frac{1}{2}$ and sending the two emerging beams to separate detectors. The observed depth of the Hong-Ou-Mandel dip \cite{HongOuPRL87} exceeded 98\%. For the subsequent measurements, the relative delay of the two photons was set at the dip minimum. The classical visibility of the interferometer was above 98\%. In order to model accurately experimental data, we developed generalized expressions for coincidence count probabilities that include effects of residual photon distinguishability, non-unit interference visibility, and fibre delivery \cite{RafalsModel}.

%\bibliography{qi}

\begin{thebibliography}{10}
\expandafter\ifx\csname natexlab\endcsname\relax\def\natexlab#1{#1}\fi
\expandafter\ifx\csname bibnamefont\endcsname\relax
  \def\bibnamefont#1{#1}\fi
\expandafter\ifx\csname bibfnamefont\endcsname\relax
  \def\bibfnamefont#1{#1}\fi
\expandafter\ifx\csname citenamefont\endcsname\relax
  \def\citenamefont#1{#1}\fi
\expandafter\ifx\csname url\endcsname\relax
  \def\url#1{\texttt{#1}}\fi
\expandafter\ifx\csname urlprefix\endcsname\relax\def\urlprefix{URL }\fi
\providecommand{\bibinfo}[2]{#2}
\providecommand{\eprint}[2][]{\url{#2}}

\bibitem{InterferometryReview}
Hariharan, P. {\em Optical Interferometry}, 2nd ed. (Elsevier, Amsterdam, 2003).

\bibitem{CavePRD81}
Caves, C. M. Quantum-mechanical noise in an interferometer. {\em Phys. Rev. D} {\bf 23}, 1693-1708 (1981).

\bibitem{GranSlusPRL87}
Grangier, P., Slusher, R. E., Yurke, B. \&  LaPorta, A.
Squeezed-light–enhanced polarization interferometer. {\em Phys. Rev. Lett.} {\bf 59}, 2153-2156 (1987).

\bibitem{XiaoWuPRL87}
Xiao, M.,  Wu, L.-A. \& Kimble, H. J. Precision measurement beyond the shot-noise limit.
{\em Phys. Rev. Lett.} {\bf 59}, 278-281 (1987).

\bibitem{HollBurnPRL93}
Holland, M. J.  \&   Burnett, K. Interferometric detection of optical phase shifts at the Heisenberg limit.
{\em Phys. Rev. Lett.} {\bf 71}, 1355-1358 (1993).

\bibitem{SandMilbPRL95}
Sanders, B. C. \&  Milburn, G. J. Optimal Quantum Measurements for Phase Estimation.
{\em Phys. Rev. Lett.} {\bf 75}, 2944-2947 (1995).

\bibitem{BollItanPRA96}
Bollinger, J. J., Itano, W. M., Wineland, D. J. \& Heinzen, D. J.
Optimal frequency measurements with maximally correlated states.
{\em Phys. Rev. A} {\bf 54}, R4649-R4652 (1996).

\bibitem{DowlPRA98}
Dowling, J. P.
Correlated input-port, matter-wave interferometer: Quantum-noise limits to the atom-laser gyroscope.
{\em Phys. Rev. A} {\bf 57}, 4736 (1998).

\bibitem{WaltPanNAT04}
Walther, P.  {\em et al.} De Broglie wavelength of a
non-local four-photon state.
{\em Nature} {\bf 429}, 158-161 (2004).

\bibitem{MitcLundNAT04}
Mitchell, M. W., Lundeen, J. S. \& Steinberg, A. M.
Super-resolving phase measurements with a multiphoton entangled state.
{\em Nature} {\bf 429}, 161-164 (2004).

\bibitem{GiovLloySCI04}
Giovanetti, V., Lloyd, S. \& Maccone, L.
Quantum-Enhanced Measurements: Beating the Standard Quantum Limit.
{\em Science} {\bf 306}, 1330 (2004).

\bibitem{HuelMaccPRL97}
Huelga, S.~F. {\em et al.}
Improvement of Frequency Standards with Quantum Entanglement.
{\em Phys. Rev. Lett.} {\bf 79}, 3865-3868 (1997).

\bibitem{RubiKausPRA07}
Rubin, M. A. \&   Kaushik, S.
Loss-induced limits to phase measurement precision with maximally entangled states.
{\em Phys. Rev. A} {\bf 75}, 053805 (2007).

\bibitem{GilbHamrJOSAB08}
Gilbert, G., Hamrick, M. \& Weinstein, Y. S.
Use of maximally entangled $N$-photon states for practical quantum interferometry.
{\em J. Opt. Soc. Am. B} {\bf 25}, 1336-1340 (2008).

\bibitem{DornDemkPRL09}
Dorner, U. {\em et al.}
Optimal Quantum Phase Estimation.
{\em Phys. Rev. Lett.} {\bf 102}, 040403 (2009).

\bibitem{BourEiblPRL05}
Bourennane, M. {\em et al.}
Decoherence-Free Quantum Information Processing with Four-Photon Entangled States.
{\em Phys. Rev. Lett.} {\bf 92}, 107901 (2004).

\bibitem{LuGaoPNAS08}
Lu, C.-Y.  {\em et al.}
Experimental quantum coding against qubit loss error.
{\em Proc. Natl. Acad. Sci. USA} {\bf 105}, 11050-11054 (2008).

\bibitem{SemiclassicalPhotodetection}
Mandel, L., Sudarshan, E. C. G.  \& and Wolf, E.
Theory of photoelectric detection of light fluctuations.
{\em Proc. Phys. Soc.} {\bf 84}, 435-444 (1964).

\bibitem{DemkDornXXX09}
Demkowicz-Dobrza\'{n}ski {\em et al.}
Quantum phase estimation with lossy interferometers.
preprint arXiv:0904.0456 (2009).

\bibitem{Helstrom1976}
Helstrom, C.~W. {\em Quantum Detection and Estimation Theory}
(Academic Press, New York, 1976).


\bibitem{BrauLanePRL92}
Braunstein, S. L.,  Lane, A. S. \&  Caves, C. M.
Maximum-likelihood analysis of multiple quantum phase measurements.
{\em Phys. Rev. Lett.} {\bf 69}, 2153-2156 (1992).

\bibitem{HongOuPRL87}
Hong, C. K., Ou, Z. Y. \& Mandel, L.
Measurement of subpicosecond time intervals between two photons by interference.
{\em Phys. Rev. Lett.} {\bf 59}, 2044-2046 (1987).

\bibitem{RariTapsPRL90}
Rarity, J. G. {\em et al.}
Two-photon interference in a Mach-Zehnder interferometer.
{\em Phys. Rev. Lett.} {\bf 65}, 1348-1351 (1990).

\bibitem{BurnWeinPRL70}
Burnham, D. C. \& Weinberg, D. L.
Observation of Simultaneity in Parametric Production of Optical Photon Pairs.
{\em Phys. Rev. Lett.} {\bf 25}, 84-87 (1970).

\bibitem{QND}
Grangier, P., Levenson, J. A. \& Poizat, J.-Ph.
Quantum non-demolition measurements in optics.
{\em Nature} {\bf 396}, 537-542 (1998).

\bibitem{RalpLangPRA02}
 Ralph, T. C., Langford, N. K.,  Bell, T. B. \& White, A. G.
Linear optical controlled-NOT gate in the coincidence basis.
{\em Phys. Rev. A} {\bf 65}, 062324 (2002).

\bibitem{OBriPrydNAT03}
O'Brien, J. L. , Pryde, G. J., White, A. G. , Ralph, T. C.
\& Branning, D.
Demonstration of an all-optical quantum controlled-NOT gate.
{\em Nature} {\bf 426}, 264-267 (2003).

\bibitem{RafalsModel}
R. Demkowicz-Dobrza\'{n}ski, unpublished.

\bibitem{MaxLik}
Kay, S. M. {\em Fundamentals of Statistical Processing, Vol. I: Estimation Theory} (Prentice Hall, Englewood Cliffs, NJ, 1993).


\bibitem{DeterministicSources}
Walmsley, I. A.
Looking to the Future of Quantum Optics.
{\em Science} {\bf 319}, 1211-1213 (2008).


\end{thebibliography}

\section*{Acknowledgements}

We wish to acknowledge insightful discussions with C. M. Caves.
This work was supported by the EU 6th Framework Programme Integrated Project Qubit Applications (contract no.\ 015848), the Polish Ministry of Science and Higher Education (grant no.\ N~N202~1489~33), the EPSRC (grant EP/C546237/1) and the Royal Society.

\end{document}